# Arithmetic logical irreversibility and the Turing halting problem

Yair Lapin, Yair.Lapin@mail.huji.ac.il

The Hebrew University of Jerusalem

Philosophy of science, philosophy of mathematics and computing

## A different approach for the halting problem of the Turing machine

March 2023 (version 5)

## Abstract:

The Turing machine halting problem can be explained by several factors, including arithmetic logic irreversibility and memory erasure, which contribute to computational uncertainty due to information loss during computation. Essentially, this means that an algorithm can only preserve information about an input, rather than generate new information. This uncertainty arises from characteristics such as arithmetic logical irreversibility, Landauer's principle, and memory erasure, which ultimately lead to a loss of information and an increase in entropy.

To measure this uncertainty and loss of information, the concept of arithmetic logical entropy can be used. The Turing machine and its equivalent, general recursive functions can be understood through the $\lambda$[1] calculus and the Turing/Church thesis. However, there are certain recursive functions that cannot be fully understood or predicted by other algorithms due to the loss of information during logical-arithmetic operations. In other words, the behaviour of these functions cannot be completely determined at every stage of the computation due to a lack of information in their definition.

While there are some cases where the behaviour of recursive functions is highly predictable, the lack of information in most cases makes it impossible for algorithms to determine if a program will halt or not. This inability to predict the outcome of the computation is the essence of the halting problem of the Turing machine. Even in cases where more information is available about the program, it is still difficult to determine with certainty if the program will halt or not.

This also highlights the importance of the Turing oracle machine, which introduces information from outside the computation to compensate for the lack of information and ultimately decide the result of the computation.

"No man ever steps in the same river twice, everything flows" Heraclitus of Efeso

---

[1] Lamda-calculus

# Introduction

This work is based on ideas arising from articles written by C. Bennet, R. Landauer, G. Chaitin.[2] The central idea of reversibility and computational entropy is based on Prof. A. Ben-Naim's explanation of entropy and the second law of thermodynamics (increased entropy).[3] In his books on these topics he suggests a more intuitive and accessible explanation of entropy and the second law based on Shannon's information theory and Boltzmann's statistical mechanics. This is a statistical mathematical explanation that can be applied to other topics because they are abstract concepts that may be disconnected from thermodynamic physics. The concepts of reversibility and entropy in this work are abstract and therefore disconnected from the physical support and discussions that arise from this.  Disputes arising from questions such as whether the interpretation of thermodynamics through information theory is correct or not correct or questions about its connection to Boltzmann's mechanics or the question of Landauer's principle that would explain Maxwell's demon paradox through the loss of[4] information at an abstract level, which many experts disagree with. These ideas of reversibility and entropy are used here as abstract tools to expose a computational limitation that explains the undecidability of the Turing halting problem.

This version of the article is different from the original versions, there are changes in the thesis. These changes are due to criticism and corrections sent to me from the original post. In this version I have tried to respond to the criticisms made to the original article. It is also an adjustment of this research to other concepts developed later, such as the conservation of information. The concepts of logical-arithmetic irreversibility and its entropy have not been criticized, they haven't been questioned; therefore, they can be considered correct. Also there haven't been any problem on the historical development of the halting problem and other computational concepts addressed in the original versions.

---

[2] In the bibliography there are the articles by these authors relevant to this work
[3] In the bibliography the first book by A. Ben-Naim about entropy and the second law
[4] the physics of the Information

But there has been criticism of the application of irreversible logic arithmetic, entropy, uncertainty, and information loss to the halting problem. Having trivial cases, in which there are primitive recursive functions that would be very predictable, that is, decidable, these functions having the same irreversible arithmetic logic and loss of information. My thesis is that in these cases is known much more about these programs or recursive functions than in many other cases where there is not enough information because they are more complex and non-trivial. The recursive functions and their complexity define hierarchies of complexity, that is, what can be calculated and what cannot be calculated by means of an algorithm; therefore, there are trivial and non-trivial cases according with the information that is having about them. So also, Bennet's reversible Turing machine susceptible to the halting problem also despite its reversibility. Being its logic-arithmetic reversible backwards but ignoring its forward behavior. Therefore, no algorithm can predict how it will behave in the future, since this information cannot be generated by another algorithm if this information is not in the machine definition. The limitation of the algorithms due to the loss of information makes it impossible to generate information that is not in the input, that is, in the program or recursive function itself.

## The Hilbert question, the theorem of Gödel and Turing's machine [5]

A famous mathematical problem and its response was what led Turing to develop his automatic calculation machine or primitive computer called Turing machine. Thus discover the problem of computable numbers that these machines have, producing or not certain types of outputs by means of automatic calculation, which would later become the famous halting problem[6]. This problem arises from a question that was formed by the famous mathematician David Hilbert in several mathematical congresses and the answer was the theorem of Gödel.  Hilbert asked the question in different ways, but it can simply be said that the question was as follows: How can it be shown that any statement or mathematical

---

question well formulated by means of a formal language of mathematical logic can be said to be true or false?   This means that you can answer to any question or mathematical statement well formulated true or false so that the math is consistent and complete and therefore you can demonstrate any theorem or affirmation. At that time, all mathematicians were convinced that any formally well-defined statement can be decided even there were unresolved mathematical problems, statements, or questions that no one knew how to solve them. But they were not unsolvable rather by a technical problem or ignorance of some detail, they were convinced that they could solve those problems in the future. No one could ever imagine at that time the possibility of another option: that there are mathematical statements that cannot be decided whether they are true or false.

The problem of decision involves finding a reliable and robust automatic method that can demonstrate all mathematical assertions using the formal language of logical inference. In response to Hilbert's question, Gödel's theorem demonstrated that mathematics cannot be both complete and consistent. However, Gödel's answer did not provide a definitive and universally accepted solution to the decision problem, and some mathematicians remained skeptical.

Alan Turing built upon Gödel's work and decisively resolved the decision problem, while also creating the theory of computability. To complete Gödel's theorem, Turing utilized the Turing machine as a tool for his demonstration.

Turing's question was somewhat different from Hilbert's one, but equivalent, this question is this: Can all numbers be calculated automatically? Are all numbers computable? This is a surprising and very rare question because no one ever saw numbers that cannot be calculated. There were not any mathematical instruments or concepts to deal with this question.  Most of the numbers people use can be calculated, but Turing suspected that there were numbers that could be formally defined but could not be calculated by finite means, like those of a person calculating on paper with a pencil. Turing defined today's well-known concept of algorithm as a process of calculation performed by a person with pen and paper. So, he wondered if an automatic machine that worked like this, like a person computing, could calculate any number? To answer this question Turing designed an abstract automatic calculation machine, a primitive computer, called a Turing machine. The idea of the machine was very simple, a person doing calculations, with limited means, step by step. The simplicity of this tool made its demonstration about incalculable numbers in a robust and out-of-all discussion result.

## The Turing machine and the automatic calculation[7]

The Turing machine is a model of computation performing an automatic, basic, and elementary calculation. Through this model you can investigate questions related to computing, computational complexity for example the number of steps that must be taken to calculate something and solve a problem, how much memory will be required for this when computer resources are limited these issues are very important. This machine can process a symbol syntax without limits in theory and can use first-order logic and arithmetic. All of this was proved by a mathematician named Alonzo Church through his theory of the $\lambda$ [8] calculus of recursive functions which is a mathematical definition equivalent to Turing's machine, but much more mathematically rigorous.[9] The $\lambda$ calculus is the mathematical foundation of computability and it allows us to view the automatic calculation machine as a series of arithmetic logical operations equivalent to a classic Turing machine. The definitions of Turing and Church are equivalent and are therefore known today by the thesis Turing/ Church. This thesis defines the concept of effective computability in mathematical logic and the concept of algorithm as an automatic calculation process. Thus, the concept of Turing equivalence allows to define every Turing machine as a system of orders more general than Turing's limited definition of the calculus machine. This is something very important and useful because it can be shown that very different computational models are equivalent to a Turing machine.  Through these new and different models, you can investigate other properties of automatic computing that are not easy to see on a classic Turing machine.

---

[7] one explanation more detailed you can find here Turing machine - Wikipedia
[8] Lambda-calculus
[9] Algorithm in the calculus $\lambda$ is based on finite recursive functions.

Finally having Turing this automatic computing model, the Turing machine he could demonstrate the decision theorem according to its equivalent approach. So, Turing represent different machines with a special number and thus flatten the differences between order tables and input states that these machines have in a way very similar to that used by Gödel in his demonstration of the indecision theorem when he codified all the formulas of formal logic. This machine numbers allowed Turing to propose that for any number that is computable must be a machine number.

In the next step Turing creates a machine that can mimic every machine, the universal machine. This machine has as input a machine number and performs the computing of that specific machine and thus checks whether the machine could be computed through the universal machine. Also, the λ calculus gives a mathematical definition of the concept of universal machine therefore this universal machine is well-founded mathematically speaking.

This work makes use of an equivalent Turing computational model in accordance with the Turing/Church thesis. This complete Turing machine is based on arithmetic logical operations that include memory erasure, something closer to a computer program with a list of arithmetic logical orders and therefore closer to the Church's λ calculus than to the classic Turing machine. But thanks to the Turing Church thesis it can be said that the conclusions affect any Turing machine or equivalent.

## The halting problem of the Turing machine as a decision problem

Decision problems are those problems on which an answer to the problem should be received in a blunt way: true or false, such as the question posed by Hilbert. You can define decision problem as any question about an infinite group of entries that must be answered if they fulfill any property, whether, therefore, indecision means that you cannot answer this question forcefully yes or no.

Turing proved through the universal machine that there are Turing machines (number of machines) that for which the universal machine cannot halt the calculation and you also cannot know if they halt one day with a result. This demonstrated that the universal machine cannot determine whether a machine halts with result by printing an output or continues to calculate in an infinite circle or freezes.[10] This fact is the halting problem of Turing machine. This is a fundamental computing calculation problem; it is directly connected to the problem of the formal math decision (Hilbert's question) and therefore the result of Turing completes the theorem of Gödel if anyone was still in doubt. Historically speaking, the concept of stop appears in Church's λ calculus as part of efficient computation, but it does not appear in the original work of Turing, who dealt with the concept of circular and non-circular machines that cannot calculate certain numbers and cannot predict what they will print. But later was done the connection between these concepts through the Turing/Church thesis[11].

For years were tried to understand more thoroughly this phenomenon that appears in different ways in formal mathematics (Theorem of Gödel) and in computing as a freeze or circularity of a program, why does this happen? and what is the cause?[12]

One of the most interesting explanations was given by one of the founders of algorithmic information theory, G. Chaitin, in an article on the theorem of Gödel and information[13], according to this article, the theorem of Gödel has an informative aspect by which any mathematical affirmation that has more information than the axiomatic system cannot be of proved using first-degree logic due to lack of information on the axiomatic basis. Therefore, the phenomenon of incompleteness proved by Gödel in the mathematical logic and Turing halting problem are common and natural phenomena, they are not rare and uncommon phenomena as they had taken them at first. This vision is at the heart of Chaitin's algorithmic information theory and the Chaitin incompleteness theorem.[14] Despite the criticisms that received these ideas,[15] for example, as how can be determined that there is more or less information on the axiomatic system, research done on random numbers and series within the framework of algorithmic information

---

theory suggests that there is some truth in Chaitin vision of things. According to this theory there is no way to calculate a random number or generate a random series because they are more informative than the finite recursive functions (algorithm) that should generate them. This is compatible with Turing's assertion about numbers that cannot be calculated with an automatic machine, but the view of algorithmic information theory is fundamentally closer to Shannon's information theory. In this work it is trying to complete the informative view of the halting problem in a different way than the theory of algorithmic information but having strong connections with it.

There are other computer scientists[16] who have thoroughly investigated this Gödel incompleteness problem and agree with the thesis about the informational aspect of the halting problem. These also point to an information limitation that occurs in all finite computing or finite theory, as the root of the problem.

According to this thesis, uncertainty is at the base of algorithmic randomness. This algorithmic randomness described by algorithmic information theory[17] is an expression of the incompleteness of arithmetic. Theorems in a finite and well-defined theory like arithmetic cannot be more complex than this theory. The most complex theorems would be improbable in theory. The information of a theory is incomplete in many cases and would be complete in trivial cases. The theory sets a limit to the complexity for the theorems that can be demonstrated within its framework, therefore, incompleteness is not accidental, and the limits imposed on computational complexity arise from the incompleteness of arithmetic and for this reason it exists halting problem.

Another very important result that is mentioned in these works and consistent with this thesis, says that the output of an arithmetic logical computation cannot have more information than the input of that computation. No more information can be obtained in the output of a program from an input by applying an arithmetic logic computation algorithm unless it is a machine that can add external information to the algorithm. Just as science needs more information to be able to explain more phenomena, a formal

---

mathematical theory needs more information to prove a theorem. We can formally express a theorem well but perhaps it is not demonstrable in the specific theory. It is something that sounds natural to us but is not nothing trivial in mathematical logic and computation.

## Recursive functions, the halting problem, and Rice's theorem

Primitive recursive functions are basic functions that are obtained by applying them a finite number of times and can be calculated by means of a computer program using, for example, a "for" type loop. A basic function can be, for example, addition or multiplication or subtraction. Applying it recursively, a primitive recursive function is obtained, such as the factorial of an integer, which is obtained by combining addition and recursive multiplication.

Most of the known computable functions are recursive primitives, therefore, they are the functions of greatest interest that can be calculated by Turing machines. But primitive recursive functions are a strict subset of general recursive ones. Church's calculus $\lambda$[18] being a formal expression of general recursive functions. The Turing Church thesis refers to the equivalence between Turing machines, effective computation, and general recursive functions. Every general recursive function can be run on a Turing machine. There are many aspects and features of computation that are studied by means of general recursive functions, and many of these features are related to primitive recursive. For example, the Ackerman function is general, but it is not primitive because it is not bounded by a finite number of steps, finiteness is a very important characteristic of primitive functions.

Primitive functions can also be total or partial. Totals would be those defined in the entire definition domain of these or partial those that are defined only in a subset of the domain. The primitive total functions are a very special subset of the partial functions. The fact that they are defined in the entire domain guarantees that any value that falls in the domain of definition of this can be calculated while in the partial ones this is not the case and therefore, they are more prone to calculation failures and to the fact that the process does not halt. For this reason, the Turing halting problem has its expression here in being able to determine if the primitive functions in a program are total or partial and this cannot be known by means of an algorithm as Rice demonstrated in his famous theorem. This theorem is central to the theory of computation and computational complexity because it establishes a very general series of problems that are undecidable by means of algorithms. This has a deep connection to the Turing halting problem as will be seen later. But it is more general and extends the classes of problems that cannot be solved by algorithms.

---

[18] Lambda-calculus

Rice's theorem establishes that all the semantic properties of recursive functions or languages recognizable by Turing machines, that is, non-trivial properties, related to their behavior, are undecidable. Trivial properties are true or false properties for all computable functions and non-trivial properties just the opposite, which cannot be established for all computable functions. In terms of general recursive functions this theorem says that there is no effective method that can establish whether a partial recursive function has some non-trivial property. There is no algorithm that can decide that a non-trivial property holds for all computable partial functions. The theorem is not applicable to properties of the Turing machine itself but to the language that this machine recognizes and the recursive function it performs, it is about the behavior of computer programs.

His proof is related to the indecision of the halting problem. If it could be established that a certain non-trivial property holds for computable partial functions by means of an algorithm, the Turing machine halting problem could be solved. But since it is not decidable, whether a program halts or not, neither can it be decided that a non-trivial property of a program holds. But this is a more general result than the Turing halting problem because it sets up a series of problems for which there is no algorithm that can decide them. For example, determine classes of recursive functions (programs) that behave in a certain way, this property being non-trivial. According to Rice's theorem, if there is at least one partial computable function in a particular class C of partial computable functions and another partial computable function that is not in C, then the problem of deciding whether a particular program computes a function in C is undecidable and this is also true for the following classes of functions[19]:

The class of recursive functions that terminates for all input.

The class of recursive functions that ends up returning 0 or not.

The class of recursive functions that produce k zeros in their calculations.

The class of recursive functions whose domain of definition is finite or not.

The class of recursive functions that are injective.

The class of recursive functions that are total or not total.

This last class describes an important undecidable problem, a partial recursive function f(x) given by a Turing T machine cannot be known if it is total or not, except in very trivial cases. The entirety of a primitive recursive function is a non-trivial property. It is also undecidable whether the condominium of a recursive function is finite or infinite. Nor can it be decided by means of algorithms if a recursive function is injective, surjective, or bijective. All these cases are not predictable by an algorithm except well-known trivial cases. These are all indecision problems derived from Rice's theorem and if they could be decided by an algorithm the halting problem would not be true.

Rice's Theorem implies that there is no algorithm that can determine, in general, whether a given recursive function computes a particular property. This is because we can code recursive functions like Turing machines, so any non-trivial property of recursive functions corresponds to a non-trivial property of languages recognized by Turing machines.

These problems are closely related to the hierarchy of arithmetic complexity. This hierarchy tells us how complex a calculus problem is in arithmetic and therefore has a lot of importance in the theory of computation. If a computational problem is isomorphic or can be reduced to one of these problems, then it has the degree of undecidability given by the complexity of this problem, and therefore this would be a difficult or impossible problem to compute.

---

[19] See A. Matos papers in the bibliography.

## The concept of arithmetic logical reversibility and memory erasure

These phenomena have been studied by C. Bennet and R. Landauer [20] in the 60s and 80s of the previous centuries on a thermodynamic physical plane with different computational devices. But they as physicist who were stayed in the thermodynamic aspects of the irreversible phenomena, less in the abstract consequences, in the logic mathematics that are the basis of computation, this area of research opened by them is known today as the physics of information. But this work is focused on the mathematical logical aspects of these concepts unrelated to physical support to give a different interpretation, thus creates some abstract concepts for these phenomena without going into the thermodynamic physical aspects. All these concepts based on mathematical logic and probability may be applied to computing, to the automatic calculation without connection to the hardware of computer or calculation machine.  Then we can discover how arithmetic logical irreversibility and memory erasure affect computation, generating computational uncertainty or improbability which would explain why the universal machine cannot calculate whether a certain Turing machine with defined input halts with a result or not. On the other hand, if we perform the calculation again the same result is received and this has an explanation, the uncertainty is local but not global, there is here a question of focus on computing or automatic calculation. The algorithm can also be treated as something potential and global such as machine code, instruction tables, program, but also as a local current process when computing. This is a macro description (program, table of orders) in front of the micro description, the states of the computing. They are not the same thing; the theory of the information revels this fact.

Landauer's principle refers to a thermodynamic aspect of computing on the physical plane. This principle sets a minimum energy limit necessary to perform an irreversible computing or calculation operation according to this when a memory bit is erased for example there is an increase in entropy and therefore power is lost by dissipation and no work can be extracted from the computer, the computer consumes power when loses information.[21]

---

[20] Below herself mentions all the Articles relevant de these authors on the subject.

Here it is suggested to interpret this in a different and more general way according to the explanation of A. Ben Naim for the rise of entropy as a general loss of information resulting from an irreversible operation at the logical, abstract level.[22] How does this happen? if we have a bit of memory and delete it , it becomes NULL and if we try to recover it, we have two options 0 or 1, each with 50% probability because there is no preference for one value or another. This creates uncertainty because it is not possible to know for sure what the bit value was therefore it is irreversible at an abstract, logical level. Many erase operations like this happen in computing as this is an allowed operation because computer resources are limited so it is necessary to delete. On the classic Turing machine, it is also a permitted operation. If we assume the possibilities that this bit could also be NULL before deletion the situation becomes even more complicated as there are 3 possibilities with 33% probability each and this generates more uncertainty.

We will discuss here the logical reversibility by means of the NAND logical gate, the famous operator NOT AND, for reasons of simplicity and its widespread use in basic digital systems. This gate or logical operator has the functional completeness property which means that you can express all the Boolean logic through combinations of this logical gate, and this is also valid for NOR, NOT OR. As a result of this property is usually based the computer's hardware that they are after all digital systems on these gates and in particular NAND because it is easier to produce them, these gates are called universal gates. But the phenomenon of logical irreversibility exists in all the Boolean logical operators that form the basis of computers and the automatic computation of recursive functions, and this will be detailed later.

---

[21] More details on this value Landauer's principle - Wikipedia
[22] In the books of Ben-Naim scored below has excellent explanations about the theme but it must be acknowledged that Bennet and Landauer they also noticed this possibility, but they still stayed in the physical interpretation, the physics of information. Here because it is computing at an algorithmic abstract interpretation is foremost.

The logical irreversibility arises from the values of the truth tables of the operator or logical gate, for example, if the output of the operator is 1 then there are 3 input options so there is uncertainty regarding the input, if it is 0 there is a single possible input then there is absolute certainty about the input of that operator.  NAND Truth Table:

| input | output |
|-------|--------|
| (1,1) | 0 |
| (1.0) | 1 |
| (0,1) | 1 |
| (0,0) | 1 |

In general, the gate or logical operator has 2 input bit and 1 bit output when in 50% of the outputs, when the value is 1, you cannot know what the exact input was because there are 3 equally likely options. This means that in the logical operator there is also uncertainty as in the case of deleting a bit because it is not possible to know with certainty all the values that were in the logical operator that have given certain results.

  More complex the logic with NAND gates generates more uncertainty by the loss of information given by the bit's differences between the inputs and outputs. I call this phenomenon logical entropy and it [23] will be analyzed later with the Shannon measure of information. Logical entropy introduces uncertainty into logical operations because you cannot know exactly what the operation was like with the inputs included and this means loss of information, at some point after many operations, you cannot know exactly how it go to a certain place or result computing and this is an irreversible situation. You could certainly know it if you save all the data from the operations done, but then the computing had to be fully computed at least once. If we add to this the deletion that it is a permitted operation on a Turing machine or equivalent machine the uncertainty for loss of information in the computation becomes very large and important.

---

[23] Using the interpretation Ben-Naim

The arithmetic reversibility works very similar to logic. Any integer that is output from an arithmetic calculation almost always has several options as input. For example, let us take the number 8 in binary notation 1000, consisting of 4 bits, being it the result of a simple arithmetic operation as addition or multiplication or a combination of these, there are several calculation possibilities as you can see and all of them possible. If we take as input two 3 bits numbers, in total 6 bits, or 3 with 3 bits numbers in total 9 bits, with the result only you cannot know which operation was made, there is a greater loss of information than there was in Boolean logic as you can see. Here are a couple of simple examples of possible operations. The operation is denoted digitally because they are performed on machines and their decimal equivalent:

011+101=1000 **(3+5=8)**, 010x100=1000 **(2x4=8)**, (010x011) +010=1000 (**(2x3) +2=8)**

010+110=1000 **(2+6=8),** 111+001=1000 **(1+7=8)**, (011x011)-001=1000 **((3x3)-1=8)**

This phenomenon is known, but they were not given enough attention and less in the context of computation or automatic calculation. As more arithmetic operations are doing in a computation, the difference between inputs and output will be larger, in these simple examples 6 bits or 9 bits inputs and 4 bits result produce a difference of 2 to 5 bits of information. Therefore, there is more uncertainty here than with the logic operation or deleting memory. Also, here so you can see there is entropy that goes up [24] with loss of information that can be calculated with the Shannon measurement of the information, but the calculation will be more complex when arithmetic operations became more complex.

---

[24] See Ben-Naim

# Information theory: Shannon measure of information, characteristics, and interpretations[25]

The Shannon measure of information is a mathematical, probabilistic tool whose function was to indicate a measure of information regardless of the content of the message. Shannon developed his theory and this measure in message communications to improve the effectiveness of communications, but this theory as it happens sometimes became something different and more extensive than the original intention. Information theory has spread to many scientific fields beyond communications engineering, sometimes with incorrect and confusing applications, but it has also given birth to very fruitful and important ideas in many areas of science. It can be said with certainty that modern digital communications are based on this theory and could not work without Shannon's ideas and her measurement of information.

Shannon's work on information began seriously in his investigation into message encryption during World War II. He was the author of a very serious mathematical research on information encoding and an analysis of message encryption in which he tried to improve the encryption of allied messages. Unlike Turing who was trying to break the encryption of messages coming out of the German encoding machine known as Enigma through of a primitive computer, Shannon tried to hide the message in a better encryption that would not destroy the content and make it unreadable. The view of each of them to the problem was therefore different, but they complement each other, they are different sides of the same coin. At that time because they were super-secret themes none of them knew which things are doing each other even they met during the war. Shannon quickly understood that a well-coded message using good encryption should look very random without losing the message itself, to permit the message decoding. This was a major limitation and the redundancies, lack of randomness, in the message was a weakness that primitive machine like the one that Turing and his team built in England could exploit to decipher these messages. Therefore, Shannon researched the structure of languages and their patterns in a new way, very different from that previously used by other experts such as linguists. He was looking for a more accurate way to tie up with messages from a mathematical point of view. He focused on redundancy patterns in messages that they were the weak point in attacking an encrypted message. This means studying the frequencies of letters, words, and their probabilities in a message. His statistical analysis revealed several important properties of a language such as that the statistical distribution of letters in each language is very different. Certain letters appear more frequently after others, specific words are more likely than others before or after certain words. Shannon calculated that the English has a redundancy of 50%, this means that you can cut a message in English by half and still have meaning and not lose the content of the message. This idea is the basis of Shannon's information theory. His work improved the compression of the probabilities of redundancies in a language, he also understood that no encryption can completely hide the message, but significant improvements can be made with techniques he discovered. This investigation into message encryption led him to the conclusion that encryption works as noise was  introduced into the message and how all these things can be calculated with some sufficiently powerful computer using statistical tools and an effective compute Turing/Church equivalent, if you could analyze the encrypted messages quickly enough could break the encryption using some kid of automatic machine, not so complex , in fact this is what Turing did in England to break the encryption of the Enigma  but Shannon did not know it.  Shannon's research proved this fact on a theoretical level and his work was classified as super-secret and it stays unpublished for many years.

Shannon's new approach was based on statistical analysis of the message totally clean of meaning. He looked at the message in a physical way, its components, symbols, words, and its relationship to information and how they appear in the message in one way and not in another. Looking at the symbols in a message according to statistical patterns Shannon observed that the information in the message is related to the appearance of those symbols in a certain way and there is uncertainty in this appearance or surprise in the order of appearance. Thus, he characterizes the statistical analysis of the information following these points:

- It is necessary to take several messages for this type of analysis, a single message is not enough.
- There are more frequent patterns than others in messages with the same information.
- The information is entropy, a concept taken from statistical mechanics (Thermodynamics) that measures the degree of order in a particle system and this is because the symbols of a message appear in a certain[26] way according to the language and theme of the message. This order is not random and therefore entropy can accurately measure the order of appearance of symbols or words.

Shannon quickly understood that her focus on the study of encrypted messages based on meaningless information can be applied to other topics such as communications. The problem with communications is that you must lower the noise that affects messages and cause errors. Therefore, this type of analysis could help improve message transmission through the communications channels and allow more messages to be passed more effectively and error free, this means strengthen the same communications channels without a physical and expensive upgrade. These things were very important to the phone company he worked after the war. Based on his information theory he suggested several improvements to communications engineers such as reinforcing redundancies in certain cases to allow message repair, message compression to better leverage communication channels, add some signal correction mechanism.[27] These recommendations based in a well-founded theory were not trivial at the time. The fundamental achievement of his theory was the use of the mathematical expression of thermodynamic entropy as a measure of the order[28] of possible configurations in a specific system as a measure of information in a message, this was really an extraordinary genius.

---

[26] the particles would be letters and the messages system.
[27] Control bit in the 8 bits byte when 7 are information and the 8th is a control bit about the previous 7, according to an accurate algorithm you can tell if the 7 bits information is correct or there is an error, some bit was changed.

The Shannon measure of the information[29] quantifies a statistical aspect of the symbols in a message as explained above and it is defined as follows: Given a group of letters or symbols in a message, usually called events or facts, given their probabilities of appearing in the message, their statistical distribution can be measured by this measure of the surprise or uncertainty or average improbability of the information in this distribution by this formula called entropy of the message according to the name Shannon gave to it. Today is tending to call it the Shannon measure of information (SMI) to differentiate it from the thermodynamic entropy in which it is inspired, this was known as Boltzmann's entropy, and it is interpreted as a measure of disorder in a system.

$$H = -\sum_{i}^{n} p_i \ \log_2 p_i$$

Where $p_i$ it represents the probability of event i, symbol i, in the message, H the entropy of the information or measure of information and it is represented in bits.

This formula is shaped like a sophisticated average over the message symbols and its original shape in statistical mechanics has a K constant in front of the sum that is related to thermodynamic units of measure. As a result of this similarity Shannon also called it entropy of information, but its intention was other, to relate the measure of the order of symbols and therefore does not mean the same thing as in thermodynamics for physical entropy that it is related to work and energy. All these things led to misinterpretations of information theory and therefore there were confusing and incorrect uses of theory.

This measure or entropy as Shannon called has 4 different and interesting interpretations, in this work I will refer to 3 of them, but also classical interpretation will be necessary to point to certain aspects of computation. It is very important to say that this measure cannot be applied to the general concept of information and less to the meaning of the information but refers to a very specific aspect of it: the existence of specific events in groups of symbols. It may use as a measure on a series of events or statistics distribution of connected events such as letters or words in a series of messages and it has no meaning to use it on a single event or unconnected events. In this work, events are the results of arithmetic logical operations and memory erasure according to the configuration of a Turing machine or computational program.

---

[28] In thermodynamics it's referred usually as disorder
[29] Known also as entropy of the information but today to differentiate it from entropy thermodynamics it is called the Shannon measure of the information.

These are the 4 interpretations of the Shannon measure of[30]information, the first 3 have probabilistic statistical roots and the fourth of a different, information-based type of how to obtain information from the statistical distribution.

1) Uncertainty in a series of experiments or events

2) Unlikelihood or Improbability of a series of events or experiments

3) Surprise or unforeseen in a series of events or experiments[31]

4) A measure of the number of bits needed to obtain accurate information about a series of data by means of binary answers to questions about this data, that is, true/false answers, these bits being 0 or 1 according to the answers.[32]

This work will mainly use interpretation 1 and 2 indicating uncertainty and unlikelihood, 4 with respect to lack of information and oracle machine and the 3 to present certain interesting characteristics of computing as a series of operations.

1) Shannon's measurement of information as a measure of the average uncertainty of a group of events

---

This interpretation arises from measure H as an average of the uncertainty of an event group, it is directly related to the statistical distribution of these group events, for example, symbols in messages. The explanation is as follows:

Given n events or experiments $x_1 x_2 \ldots x_n$ whose probabilities in distribution are $p_1 p_2 \ldots p_n$, when p=1 the event is 100% certain but for p < 1 it is less certain and more uncertain. If $-\log_2 p_i$ it is big, p is small. The -log strongly decreasing when the probability rises therefore $p_i \log_2 p_i$ indicates uncertainty regarding the i event because the possibility of the event is small and the sum of all raises them, this sum basically indicates the average uncertainty of all events according to the statistical distribution of them.

Graph 1 shows the -log p function strongly decreasing and not linearly with the increase in the probability of the event and goes up exponentially when in event it becomes unlikely.

The -log p and -p log p graphs will help for better understanding of the first two statistical interpretations of the Shannon measure of information as uncertainty and improbability. Close to probability 1 of the event in graph 2 you see that -p log p is close to 0 and therefore this indicates uncertainty regarding the event and just the maximum certainty would be at the probability 0.3679 when the maximum of -p log p is 0.53, this behavior permeates all probabilistic interpretations of the Shannon measure.

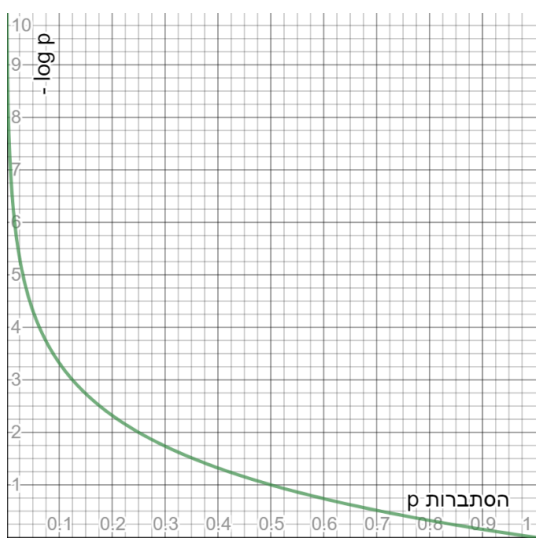 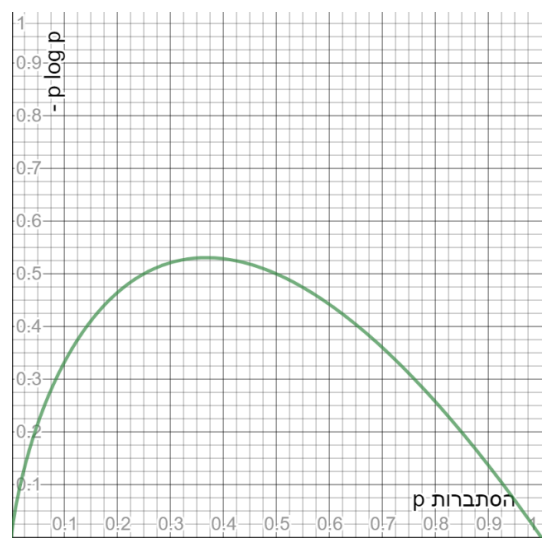

**Graph 1**          **Graph 2**

## 2) The unlikelihood or lack of expectation or expectedness average of a group of events

This interpretation of H is like the previous one but in terms of expectation and possibility of an event in a series of connected events and is also probabilistically based. When p is very small it is very feasible that the event does not happen and when approaching 1 it is very feasible to happen. So being -log p a strongly decreasing function, the situation is reversed because it balances to the probability of the p event and therefore the multiplication of these, $p_i \log_2 p_i$ the function -log p indicates improbability and lack of expectation when p event expectation, but -log p decreases strongly to 0 and therefore H indicates with the sum the average improbability or lack of expectation of these events.

### 2) The surprise or unforeseen average in a group of events

This is the classic interpretation of the measure, the one that Shannon referred to extensively in his work and which was quoted above because it is strongly connected to the problem of encryption and message transmission focusing on the redundancy patterns of a message. When talking about surprise and lack of expectations, it refers to a serial current of symbols, each new symbol is a surprise, you cannot know what the next symbol will be, but you can calculate your probability in the current according to the previous occurrences. This interpretation helps clarify Shannon's ideas when developing information theory and its measurement. It is very difficult to understand the information theory without understanding this.

## 4) Information measuring at several events

This interpretation is somewhat more sophisticated than the previous ones and is not related to the probability of the event. It should be remembered that this measure is not information, but you can get information from this, information regarding a statistical distribution of events, from a group of data that have well-defined probabilities.

This quantitative interpretation of H entropy is directly related to the lack of information produced by uncertainty in a series of statistical data and how information can be obtained from this statistical distribution in a minimum number of steps. This point is very important for the discussion below regarding obtaining information under conditions of uncertainty. To explain this interpretation, we will define a game, in which we have a table divided into 8 lockers, these lockers divided by a somewhat higher wall to prevent an object thrown to the table from being on the edge and not entering a locker.[33] The 8 lockers are numbered from 1 to 8. In this game there are 2 players, and one randomly throws a coin on the table and the other must guess in what locker it is in and this by asking a minimum of binary questions, namely questions whose answer is yes or no. If the player asks without any order or strategy, he/she can guess the first one, but also easily may get to the maximum of 8 questions then the average of questions of the player who handles so will be high. The player who wants to win must play with some strategy to keep an average of questions low on all laps of this[34] game. In this game players face a problem of uncertainty, and it is about getting the exact information in an optimized way through binary questions, there is a strategy that will allow the player to use it to get to the exact place of the coin in a minimum of steps. Without this strategy the average steps will be higher and there is a mathematical prove for this, but we will not enter this prove here.

The following graph depicts the game, the table divided into 8 lockers, the coin that is thrown in a random way and which the other player must guess where it is.

| | | | |
|---|---|---|---|
| | | | 🔵 |
| | | | |

The correct strategy is to split the table in 2 halves, two areas, right and left and ask if it is the coin on the left or right and so again on the area, we get true answer until you reach the locker, this strategy limit the possibilities all the time by converging towards the locker where the coin is located. This is the most effective way to find the place where the coin is located.[35] It is said that the entropy H, measure Shannon of the information, indicates exactly the number of steps that must be done on average to reach the exact position of the coin, in this case the probability of the event is and $\frac{1}{8}$ then calculating the entropy for 8 lockers we get the following:

$$-\sum_1^8 \frac{1}{8} \text{ x } log_2 \frac{1}{8} = 3 \text{ bits}$$

This means that using the right strategy in 3 questions by average we got to the place of the coin, and this would be the most effective way to get this information and anyone who does not use this strategy will reach higher averages of questions.

If the probability of the event is not $\frac{1}{8}$ as in this simple example due to different lockers size, the number of questions will change according to the calculation of H but the basic idea would be the same, divide the table into areas with the highest probability, here because they are equally likely events is divided into areas with 50% probability at each step but if there is any asymmetry for having higher probabilities you should try to take areas with the maximum probability. All of this depends on the distribution of probabilities of events.

This interpretation allows us to obtain information from a series of data about which there is a priori uncertainty, but we are interested in a posteriori information through a convergent process of limitation of possibilities strategy that minimizes the number of steps or questions to obtain the information. Attention, the answer to each binary question is one bit of information and therefore in this case 3 bits of average information are necessary to locate the coin in the locker.

---

[35] Ben-Naim, information theory, chapter 2, properties of the Shannon measure of the information

# Arithmetic logical entropy and erasure of information in terms of information theory

If we apply the Shannon measure on the NAND logical operator, for example, but this is also true for any other logical operator and calculate their entropy we will get their uncertainty. The results of this logical operator are given by its truth table and therefore there are 2 possible values 1 or 0, with 50% of probability to each one, if we take the values in the input and output table the situation is as follows: for (1.1) -> 0, 100% input/output certainty, probability 1. So, the probability of event 0 is according to conditional probability, Bayes, 1x $\frac{1}{2}$ this is $\frac{1}{2}$ $and$ $this$ represents half of the logical operator's possibilities.

For the output 1 is more complicated because there are multiple inputs for the same result (0.1), (0.0),(1,0) ->1 , the input has a $\frac{1}{3}$ of probability each one and the conditional probability of the events is then $\frac{1}{6} = \frac{1}{2} x \frac{1}{3}$ entropy for this logical operator is:

$$H= - [ 3 \times ( \frac{1}{6} \ x \ log_2 \frac{1}{6} ) + ( \frac{1}{2} \ x \ log_2 \frac{1}{2} ) \ ] = 1,7924 \text{ bits}$$

This is almost 1.8 bits namely 1.8-bit uncertainty approximately per simple logical operation and according to interpretation 4 you must ask 1.8 minimum binary questions to get the values of the input of the logical operator and reconstruct the operation completely, inputs and outputs. In event 1 you can ask if the first entry is 0 or 1 and thus reduce the possibilities, in the worst case you will make 3 steps in the best 1 and thus give the average about 2.

In memory erasure entropy is 1, namely 1 bit uncertainty, in case that the bit had two chances 1 or 0 and therefore it is not known what there was in the bit before the deletion. But if there is one more possibility, NULL namely empty, there are 3 possible states before erasing, this empty bit chance cause to the entropy goes up even higher because uncertainty goes up. Basic case without empty is:

$$H= - 2 \times ( \frac{1}{2} x \ log_2 \frac{1}{2} ) = 1 \text{ bit}$$

Case with NULL as a possible event

$$H = -3x \left( \frac{1}{3}x \, log_2 \frac{1}{3} \right) = 1.58 \text{ bit}$$

In elementary arithmetic or basic calculations can be easily seen as entropy goes up when calculations become more complex, inputs are larger with the combination of more operations. A simple but illustrative example, 011+101=1000 **(3+5=8)** shown above. Here there are 6 bits input representing integers, each bit has 2 options 1 or 0 then the probability in each bit input is $\frac{1}{2}$ when we have a 4-bit output, its entropy is calculated as well:

$$H = -6 \times \left( \frac{1}{2}x \, log_2 \frac{1}{2} \right) = 3 \text{ bits}$$

This means 3 bits of uncertainty for this simple arithmetic operation. Arithmetic entropy will rise rapidly for more complex operations.

If we apply here the Shannon's interpretation 4, we will see in this example a couple of interesting things. If we ask the binary questions to obtain information, each answer would be a bit of information, having the output of the calculation plus these bits we will be able to completely reconstruct the operation then the entropy also represents the bits or information lost in the arithmetic operation. In the logical output bit operator + 1.8-bit information fully describes an operation with 3-bit input/output namely 3 bits is not required. In the deletion certainly 1 bit is enough to know if there were 1 or 0 only. In the simple arithmetic of this example 4 output bits plus 3 bits are sufficient to reconstruct the operation, these 3 bits could be one of the inputs and the subtraction operation on the output we get the other input from the original operation. This is an important fact because these bits given by the entropy would be the information lost in an irreversible operation and would make it reversible if these bits were obtained. This means that they complete the output of the operation, and the entire input is not required to be a reversible operation according to these simple examples. In machines or programs where millions of logical-mathematical operations are performed and erasing memory the information losses during computing would be huge.

## Information loss due to arithmetic logic entropy and memory erasure

Due to the loss of information in the computation in complex programs, several steps forward and even several steps backward are unknown, except in cases such as Bennet's reversible machine. Therefore, in most cases it's not possible to predict exactly what will happen in these programs several steps forward or backward, except in trivial cases. In Rice's terms, it's not known exactly what non-trivial properties there are in that program or recursive function based on its definition alone. More information would be needed to determine it, and an algorithm would not suffice to determine this because algorithms do not generate more information on their inputs. In trivial cases, for recursive functions or programs, would know where their behavior would be limited in some way. This would be extra information that is not always available, for example, knowing the definition domain / image well or whether they are finite or not, or other semantic information about them that can add more certainty to the computation.

Given these conditions, the Turing result on the universal machine is very clear and reasonable. Therefore, cannot be calculated by means of an algorithm if a Turing machine (computer program) would halt or not with output. This algorithm limitation would be a consequence of the uncertainty of the computation and its loss of information. The halting conditions of a Turing machine or program cannot be calculated in a deterministic way due to computational uncertainty and lack of trivial information about it because this information cannot be generated with another algorithm.

## A different interpretation for the Oracle Turing machine[36]

The oracle Turing's machine is a common machine that is connected to an oracle that can answer questions sent by the machine immediately, yes, or no, without any computing. In this context this machine can face decision problems because it does not have to perform any calculation but ask to the oracle, this type of machine can therefore solve the halting problem and thus do things that the common machine cannot calculate automatically.[37] It generally does not matter how the oracle works, but this can be a giant database, for example. In the configuration of the oracle machine there is an order called ASK which is the call to the oracle and ask something for example whether such a Turing machine that is running now ends with a printed response and therefore a universal oracle machine could theoretically solve the halting problem.

---

[36] one explanation more detailed information about this type of machine here Oracle machine - Wikipedia

The oracle machine is an open system because it can receive information from outside the calculation or computing unlike a common machine that it is closed system, where there is only one input and then the computation process. The machine oracle enters information from outside to complete missing information and make an incalculable decision otherwise. The oracle generates more information that is not possible by computing using finite recursive functions.[38] The oracle would complete the missing information, and this would be the informative aspect that Chaitin was talking about in his article mentioned here and in general in his theory.

On an oracle machine, the output can be more informative than the input. Because information is being entered into the computation and not only calculating. This would be the big difference with respect to a common machine.

## Conclusions

The loss of information due to logical or arithmetic operations or memory erasure is the root of computational uncertainty or improbability. This would limit the possibility of an algorithm to generate more information about an input, at most it would preserve this information if the loss of information is neutralized by saving the whole computation. But no more information is generated than there is.

---

[37] Through a algorithm
[38] algorithm

Turing's result is very clearly proven and the generalization of it by Rice's theorem extends this problem to more cases. The semantic properties of a program are not calculable due to lack of information. This agrees with the view of many experts mentioned here regarding the importance of information in understanding the halting problem and Gödel's theorem. An algorithm defines a rule, but in many cases, it does not define all the potential data of the rule. These data would be a surprise in terms of Shannon information, therefore not computable by other algorithms that are also limited.

Here I must mention Hintikka's famous philosophical comment that says that Gödel's proof of incompleteness has no importance from a philosophical point of view since it does not explain the elementary reasons why arithmetic is incomplete, it only proves that it is. In this work, as well as in other works mentioned here, an attempt has been made to present a probable hypothesis as to why incompleteness exists in arithmetic and why the Turing halting problem exists. The hypothesis here is that the arithmetic logic and its entropy create an information barrier about the system. A limit of the probable in a system or theory, more information cannot be generated based on a given information. This would affect the forward and backward computation. But this hypothesis has not yet been fully investigated and formalized enough to give a definitive answer to the philosophical question that Hintikka leaves us. It would be necessary to further investigate the arithmetic logical entropy and the loss of information, the uncertainty generated by this, to give a definitive answer to Hintikka's fundamental question.

## Acknowledgements: My thanks to Prof. A. Ben-Naim for his explanations and comments on information theory and its connection to statistical mechanics and thermodynamics. My thanks to Gregory Chaitin for his opinion on the ideas of the article and recommendations, I also appreciate the help of Cristian Calude for his opinion, recommendations and corrections that have been very useful. I also thank my old friend from the Technion, Ing. Marcelo Yoffe from Intel Israel, whose comments and recommendations have been essential to write this article. I also thank the people who have shown interest in this thesis and have read the drafts posted on the internet, for their comments, critics, corrections, and interest in the subject.